\documentclass[aps,prd,10pt,showpacs,amsmath,amssymb,twocolumn,nofootinbib,nobibnotes,preprintnumbers
]{revtex4-2}

\usepackage{float}
\usepackage{mathrsfs,amsfonts}
\usepackage{mathtools}
\usepackage{tensor}
\usepackage[T1]{fontenc} 
\usepackage{xcolor}
\usepackage{url}
\usepackage[hyperindex,breaklinks]{hyperref}
\usepackage{graphicx}

\pdfoutput=1
\newcommand{\sig}{\ensuremath{{\hat{\sigma}}}}
\newcommand{\s}{\ensuremath{{\hat{s}}}}

\begin{document}
\allowdisplaybreaks[1]
\title{Scalaron-Higgs inflation reloaded: Higgs-dependent scalaron mass and primordial black hole dark matter}
\author{Anirudh Gundhi~${}^{a,b}$}
\author{Christian F. Steinwachs~${}^{c}$}
\email{anirudh.gundhi@phd.units.it}
\email{christian.steinwachs@physik.uni-freiburg.de}
\affiliation{${}^{a}$Dipartimento di Fisica, Universit\`a degli Studi di Trieste, Strada Costiera 11, 34151 Miramare-Trieste, Italy\\
${}^{b}$ Istituto Nazionale di Fisica Nucleare, Trieste Section, Via Valerio 2, 34127 Trieste, Italy}
\affiliation{${}^{c}$Physikalisches Institut, Albert-Ludwigs-Universit\"at Freiburg,\\
Hermann-Herder-Str.~3, 79104 Freiburg, Germany}
%
%
\begin{abstract}
We propose an extension of the scalaron-Higgs model by a non-minimal coupling of the Standard Model Higgs boson to the quadratic Ricci scalar resulting in a Higgs-dependent scalaron mass. The model predicts a successful stage of effective single-field Starobinsky inflation. It features a multi-field amplification mechanism leading to a peak in the inflationary power spectrum at small wavelengths which enhances the production of primordial black holes.
The extended scalaron-Higgs model unifies inflationary cosmology with elementary particle physics and explains the origin of cold dark matter in terms of primordial black holes without assuming any new particles.  
\end{abstract}
%
%
\maketitle
%
%
\section{Introduction}
\label{Sec:Introduction}
In view of the countless number of inflationary models, predictability as well as theoretical motivation become even more important. In this respect, Starobinsky's quadratic $f(R)$ model and the model of Higgs inflation stand out.
Starobinsky's model is the natural geometric extension of Einstein's theory in which the higher derivatives in the quadratic curvature scalar lead to the emergence of an additional massive scalar propagating degree of freedom, the scalaron, driving inflation \cite{Starobinsky:1980te}.
In contrast, in Higgs inflation, the inflaton is identified with the Standard Model (SM) Higgs boson non-minimally coupled to the Ricci scalar \cite{Bezrukov2008}. Quantum corrections dominated by the heavy SM particles \cite{Barvinsky2008} establish the connection between particle physics at the electroweak scale and cosmology at the energy scale of inflation via the 
renormalization group (RG) running \cite{Bezrukov2009a,DeSimone2009,Barvinsky2009,Bezrukov2011,Barvinsky2012}, see \cite{Steinwachs:2019hdr} for a recent review on the Higgs field in cosmology.  
Both models lead to the same predictions for the spectral inflationary observables \cite{Barvinsky2008,Bezrukov2012,Kehagias2014}. This is a particular manifestation of a more general equivalence between $f(R)$ theories and scalar-tensor theories for different field parametrizations at the classical and quantum level \cite{Steinwachs2014,Kamenshchik2015,Ruf2018a}. 

The unification of these two models results in the two-field scalaron-Higgs model \cite{Ema2017a,Wang2017,He:2018gyf,Gundhi:2018wyz} (in the Palatini formalulation of the model \cite{Antoniadis:2018yfq,Gialamas:2019nly,Tenkanen:2019jiq,Tenkanen:2020dge,Gialamas:2020snr} the scalaron is not a dynamical degree of freedom).  
Various aspects of this model have been studied, including the dependence of the inflationary dynamics on the initial conditions \cite{Gundhi:2018wyz,Enckell:2018uic,Tenkanen:2020cvw}, the properties of its RG improvement \cite{Salvio2015,Calmet:2016fsr,Gorbunov2018,Ghilencea2018,Gundhi:2018wyz,Samart:2018mow,Cheong:2019vzl,Ema:2020evi}, the stabilization of the SM vacuum \cite{Ema2017,Gundhi:2018wyz}, the (p)reheating scenario \cite{He:2018mgb,Bezrukov:2019ylq,He:2020ivk,Bezrukov:2020txg}, and a multi-field amplification mechanism that leads to features in the inflationary power spectrum \cite{Gundhi:2018wyz}.
Another interesting class of multi-field extensions of Higgs inflation and Starobinsky's $R+R^2$ model is based on a coupling to a dilaton field \cite{Shaposhnikov:2008xi,GarciaBellido:2011de, Blas2011b, Kaneda2016, Gundhi:2020kzm}. 

In contrast to single-field models of inflation \cite{Garcia-Bellido:2017mdw,Kannike:2017bxn,Ezquiaga:2017fvi,Motohashi:2017kbs,Rasanen:2018fom,Mishra:2019pzq}, multi-field models offer additional mechanisms for the amplification of the adiabatic power spectrum due to the multidimensional potential landscape and the curved field space geometry. 
In particular, such an amplification can result in the formation of peaks in the adiabatic power spectrum at small wavelengths. For peaks with sufficiently large amplitudes, the production of Primordial Black Holes (PBHs) is greatly enhanced and might explain the observed Cold Dark Matter (CDM) content of the Universe, see \cite{M3,Sasaki:2018dmp,Carr:2020xqk} for a review. The production of PBHs in multi-field inflation has recently been realized in a variety of models, see e.g.~\cite{Pi:2017gih,Inomata:2017vxo,Canko:2019mud,Palma:2020ejf,Fumagalli:2020adf,Braglia:2020eai,Aldabergenov:2020bpt,Gundhi:2020kzm}.
The formation of PBHs by an effective single-field ultra-slow roll mechanism has also been investigated in a fine-tuned RG-analysis of the scalaron-Higgs model \cite{Cheong:2019vzl}. 
Aside from offering an explanation for the origin of the observed CDM, PBHs offer a unique opportunity to constrain the adiabatic power spectrum at small wavelengths not accessible to Cosmic Microwave Background (CMB) measurements. 

In this article, we extend the scalaron-Higgs model by a Higgs-dependent scalaron mass crucial for the successful realization of the multifield ``isocurvature pumping'' amplification mechanism described in \cite{Gundhi:2018wyz,Gundhi:2020kzm}. This results in the formation of a single peak in the adiabatic power spectrum triggering the formation of PBHs at small wavelengths. We find simple scaling relations among the model parameters, which allow to adjust the amplitude and the location of the peak in the power spectrum, such that a maximum total PBH-CDM mass fraction can be realized in all observationally viable mass windows. 
  
The model proposed in this article provides a unified description of inflationary cosmology, CDM, and elementary particle physics without assuming any new physics except for a non-minimal coupling of the SM Higgs boson to the modified gravitational $R+R^2$ sector.    
%
%
\section{Extended scalaron-Higgs model}
\label{Model}
We assume the SM embedded into curved spacetime with a modified graviton-Higgs sector given by
\begin{align}\label{JFact}
S_{\mathrm{gh}}=\int\mathrm{d}^4x\sqrt{-g}\left[f(R,\varphi) 
-\frac{1}{2}g^{\mu\nu}\partial_{\mu}\varphi\partial_{\nu}\varphi\right].
\end{align}
The curvature-Higgs-dependent function in \eqref{JFact} reads
\begin{align}\label{ffunction}
f(R,\varphi) = 
 \frac{U(\varphi)}{2}\left(R + \frac{1}{6\,M^2(\varphi)}R^2\right)-V(\varphi).
\end{align}
The structure of the action \eqref{JFact} with the function \eqref{ffunction} has been investigated in \cite{Gundhi:2020kzm} in the context of an abstract dilaton field $\varphi$. In contrast, in this article, we study the implications of identifying $\varphi$ with the SM Higgs field. This model provides a natural extension of the scalaron-Higgs model considered in \cite{Ema2017a,Wang2017,He:2018gyf,Gundhi:2018wyz}.
We formulate the model as a two-field scalar-tensor theory.\footnote{The (on-shell) equivalence between $f(R)$ theories and scalar-tensor theories and different field parametrization at the classical and quantum level has been demonstrated in \cite{Kamenshchik2015,Ruf2018a}.} The auxiliary field $\chi$ emerges when formulating the $f(R,\varphi)$ theory as a two-field scalar-tensor theory \cite{Gundhi:2018wyz}. Transforming in addition to the the Einstein frame (EF) by performing the non-linear field redefinitions 
\begin{align}
g_{\mu\nu}=\frac{1}{2}\,\frac{M_{\mathrm{P}}^2}{\chi^2}\hat{g}_{\mu\nu},\qquad
\chi=\frac{M_{\mathrm{P}}}{\sqrt{2}}\exp\left(\frac{\hat{\chi}}{\sqrt{6}M_{\mathrm{P}}}\right),\label{conftraf}
\end{align}
the action \eqref{JFact} is written as two-field action in the EF
\begin{align}
S[\hat{g},\Phi]={}
\int\mathrm{d}^4 x\sqrt{-\hat{g}}&\left[\frac{M_{\mathrm{P}}^2}{2}\hat{R}-\frac{\hat{g}^{\mu\nu}}{2}G_{IJ}\Phi^{I}_{,\mu}\Phi^{J}_{,\nu}-\hat{W}\right].\label{EFact}
\end{align}
Here, $\hat{\chi}$ is the scalaron, effectively emerging from the higher derivatives present in $R^2$ term in \eqref{ffunction}. A more detailed presentation of the transition from \eqref{JFact} to \eqref{EFact} can be found in \cite{Gundhi:2018wyz}.
The local scalar field coordinates $\Phi^{I}(x)$ and the metric $G_{IJ}$ on the scalar field-space manifold are defined by
\begin{align}
\Phi^{I}=\left(\begin{array}{c}\hat{\chi}\\\varphi\end{array}\right),\qquad G_{IJ}(\Phi)=\left(
\begin{array}{cc}
1&0\\
0&F^{-1}\left(\hat{\chi}\right)
\end{array}
\right)\,. \label{GMetric}
\end{align}
The scalar two-field potential $\hat{W}(\Phi)$ in the EF reads
\begin{align} \label{EFpotW}
\hat{W}(\varphi,\hat{\chi}) = \frac{V}{F^{2}}+ \frac{3}{4} m^2\, M^2_{\mathrm{P}} \left( 1 -\frac{U}{M^2_{\mathrm{P}} F}\right)^2.
\end{align}
We have introduced the parametrization
\begin{align}
F (\hat{\chi}):={}& \exp\left( \sqrt{\frac{2}{3}} \frac{\hat{\chi}}{M_{\mathrm{P}}}\right),\label{Fchi}\\ m^2(\varphi):={}&M^2(\varphi)\frac{M_{\mathrm{P}}^2}{U(\varphi)}.\label{mphi}
\end{align}
The extended scalaron-Higgs model is defined by 
\begin{align}
U(\varphi)&= M^2_{\mathrm{P}}+\xi\varphi^2, \label{f1}\\
m^2(\varphi)&= m_0^2+\zeta\varphi^2, \label{f2}\\ 
V(\varphi)  &=\frac{\lambda}{4}\left(\varphi^2-\nu^2\right)^2. \label{f3}
\end{align}
Here, $U(\varphi)$ corresponds to an effective Higgs-dependent gravitational constant with the reduced Planck mass ${M_{\rm P}=1/\sqrt{8\pi G_{\mathrm{N}}}\approx 2.4\times 10^{18}\;\mathrm{GeV}}$ (in natural units ${c=\hbar=1}$) and the non-minimal coupling $\xi$, while $M(\varphi)$ leads to an effective Higgs-dependent scalaron mass $m(\varphi)$ with the constant scalaron mass $m_0$ and non-minimal coupling $\zeta$. The potential $V(\varphi)$ is defined by the SM Higgs potential with the quartic self-coupling $\lambda$ and the symmetry breaking scale $\nu\approx246\;\mathrm{GeV}$.
The explicit form of the functions in \eqref{f1}-\eqref{f3} may also be motivated by the their limits for small and large values of the Higgs field. In the limit $\varphi/M_{\mathrm{P}}\to0$, the original Starobinsky model \cite{Starobinsky:1980te} is recovered, while for $\varphi/M_{\mathrm{P}}\to\infty$, the model features an asymptotic scale invariance.
The main generalization compared to the scalaron-Higgs model investigated in \cite{Gundhi:2018wyz} is the Higgs-dependent function \eqref{f2}. In view of $\nu/M_{\mathrm{P}}\approx10^{-16}$, we neglect the constant $\nu$ for the inflationary analysis. Under these assumptions, the inflationary EF two-field potential \eqref{EFpotW} reduces to
\begin{align}\label{EFpotWExpl}
\hat{W}(\hat{\chi},\varphi) = \frac{\lambda\varphi^4+ 3 M^2_{\mathrm{P}}\left(m_0^2+\zeta\varphi^2\right) \left( 1+ \xi\frac{\varphi^2}{M^2_{\mathrm{P}}}-F\right)^2}{4F^2}. 
\end{align}

\section{Covariant multi-field formalism} 
Following the general treatment in \cite{Gundhi:2018wyz}, we formulate the inflationary dynamics of the background and the perturbations in terms of the covariant multi-field formalism.
The line element of the perturbed flat Friedmann-Lema\^itre-Robertson-Walker (FLRW) universe reads 
\begin{align}
\mathrm{d}s^2=&-\left(1+2A\right)\mathrm{d}t^2+2a B_{,i}\mathrm{d}x^i\mathrm{d}t\nonumber
\\&+a^2\left(\delta_{ij}+2E_{ij}\right)\mathrm{d}x^i\mathrm{d}x^{j}.
\end{align}
Here, $t$ is the cosmic time, $a(t)$ is the scale factor, $i,\,j,\ldots=1,2,3$ are spatial indices, ${\delta_{ij}=\mathrm{diag}(1,1,1)}$ is the flat spatial metric and ${E_{ij}:=\psi\delta_{ij}+E_{,ij}}$. The scalar metric perturbations $A(t,\mathbf{x})$, $B(t,\mathbf{x})$, $\psi(t,\mathbf{x})$, and $E(t,\mathbf{x})$ combine with the scalar perturbations $\delta\Phi^{I}(t,\mathbf{x})$.
The Friedmann equations and the Klein-Gordon equations for the homogeneous scalar field multiplet $\Phi^{I}(t)$ read
\begin{align}
H^2 ={}& \frac{1}{3M^{2}_{\mathrm{P}}}\left[\frac{1}{2}G_{IJ}\dot{\Phi}^I\dot{\Phi}^J + \hat{W}(\Phi)\right],\label{Friedmann1}\\
\dot{H} ={}& -\frac{1}{2M^{2}_{\mathrm{P}}}G_{IJ}\dot{\Phi}^I\dot{\Phi}^J,\label{Friedmann2}\\
D_{t}\dot{\Phi}^I={}& - 3H\dot{\Phi}^I - G^{IJ}\hat{W},_{J}.\label{KleinGordon}
\end{align}
The dot is shorthanded for $\partial_t$. The Hubble parameter $H(t)$ and the covariant time derivative $D_t$ are defined by
\begin{align}
H(t):=\frac{\dot{a}(t)}{a(t)},\qquad D_{t}V^{I}:=\dot{V}^I+\dot{\Phi}^{J}\Gamma^{I}_{JK}(\Phi)V^{K}.
\end{align}
The connection $\Gamma^{I}_{JK}$ is defined with respect to \eqref{GMetric} and the unit vector along the background trajectory reads
\begin{align}
\hat{\sigma}^I={}& \frac{\dot{\Phi}^I}{\dot{\sigma}},\qquad\dot{\sigma}= \sqrt{G_{IJ}\dot{\Phi}^I\dot{\Phi}^J},\qquad G_{IJ}\hat{\sigma}^{I}\hat{\sigma}^{J}={}1.\label{dsigma}
\end{align}
The unit vector $\hat{s}^{I}$ orthogonal to $\hat{\sigma}^{I}$ satisfies
\begin{align}
\label{norms}
G_{IJ}\hat{s}^I\hat{s}^J = 1,\qquad G_{IJ}\hat{s}^I\sig^J = 0.
\end{align} 
The unit vector $\hat{s}^{I}$ is proportional to the acceleration vector $\omega^I$ which defines the turn rate $\omega$,
\begin{align}
\omega^{I}=D_{t}\hat{\sigma}^{I},\qquad \omega={}\sqrt{G_{IJ}\omega^I\omega^J},\qquad \hat{s}^{I}=\frac{\omega^{I}}{\omega}.\label{TurnVector}
\end{align}
Instead of the perturbations $\delta\Phi^{I}(t,\mathbf{x})$, we work with the gauge-invariant Mukhanov-Sasaki variable \cite{Mukhanov1988,Sasaki1986,Greenwood2013},
\begin{align}
\delta\Phi^{I}_{\mathrm{g}}=\delta\Phi^I+\frac{\dot{\Phi}^{I}}{H}\psi.\label{MukSasPhi}
\end{align}
The equation for the Fourier modes of the perturbation $\delta\Phi^{I}_{\mathrm{g}}(t,\mathbf{k})$ is found to be \cite{Sasaki1996,Nakamura1996,Greenwood2013},
\begin{align}
D^2_t\delta\Phi^I_\mathrm{g}+ 3HD_t \delta\Phi^I_\mathrm{g}
+\left(\frac{k^2}{a^2}\delta^I_J+\tensor{\Omega}{^{I}_{J}}\right)\delta\Phi^J_\mathrm{g}=0.\label{DynEQPertPhi}
\end{align}
Following the conventions introduced in \cite{Gundhi:2018wyz}, $\tensor{\Omega}{^{I}_{J}}$ and the effective mass tensor $\tensor{M}{^{I}_{J}}$ are defined by
\begin{align}
\tensor{\Omega}{^{I}_{J}}&={}\tensor{M}{^I_J}-M^{-2}_\mathrm{P}a^{-3}D_t\left(\frac{a^3}{H}{\dot{\Phi}}^I{\dot{\Phi}}_J\right),\label{Omega}\\
M_{IJ} &={} \nabla_I\nabla_J\hat{W} + R_{IKJL}\dot{\Phi}^K\dot{\Phi}^L.\label{EffMassMatrix}
\end{align}
Here $R_{IJKL}$ is the Riemannian curvature tensor associated with the curved scalar field space manifold.
Projecting \eqref{MukSasPhi} along $\hat{\sigma}^{I}$ and $\hat{s}^{I}$ defines the adiabatic and isocurvature perturbations
\begin{align}\label{QSigma}
Q_{\sigma}=\hat{\sigma}^{I}G_{IJ}\delta\Phi^{J}_{\mathrm{g}},\qquad Q_{\mathrm{s}}=\hat{s}^{I}G_{IJ}\delta\Phi^{J}_{\mathrm{g}}.
\end{align}
Inserting  $\delta\Phi_{\mathrm{g}}^{I}=Q_{\sigma}\hat{\sigma}^{I}+Q_{\mathrm{s}}\hat{s}^{I}$
into \eqref{DynEQPertPhi}, the dynamical equations for the Fourier modes $Q_{\sigma}(t,\mathbf{k})$ and $Q_{\mathrm{s}}(t,\mathbf{k})$ in the large wavelength limit $k\ll aH$ read
\begin{align}
\ddot{Q}_{\sigma} + 3H\dot{Q}_{\sigma} + \Omega_{\sigma\sigma}Q_{\sigma}={}& f(\mathrm{d}/\mathrm{d}t)(\omega Q_{\mathrm{s}}),\label{EQPertQsig} \\
\ddot{Q}_\mathrm{s} + 3H\dot{Q}_\mathrm{s} + m_\mathrm{s}^2Q_\mathrm{s}={}&0\label{EQPertQs}.
\end{align}
The effective masses $\Omega_{\sigma\sigma}$ and $m_{\mathrm{s}}^2$ are defined by projecting \eqref{Omega} and \eqref{EffMassMatrix} respectively and include additional contributions of the turn rate
\begin{align}
\Omega_{\sigma\sigma}=\hat{\sigma}^I\hat{\sigma}^J\Omega_{IJ}-\omega^2,\qquad
m^2_{\mathrm{s}}= \s^I\s^J M_{IJ} + 3\omega^2\label{IsoMassDef}.
\end{align}
The operator $f(\mathrm{d}/\mathrm{d}t)$ in \eqref{EQPertQsig} is defined by
\begin{align}
f(\mathrm{d}/\mathrm{d}t)=2\left[\frac{\mathrm{d}}{\mathrm{d}t}-\left(\frac{W,_{\sigma}}{\dot{\sigma}} + \frac{\dot{H}}{H}\right)\right].
\end{align}
Only if the combination of $\omega$ and $Q_{\mathrm{s}}$ is sufficiently large, $Q_{\sigma}$ is sourced by the ``isocurvature pumping'' mechanism discussed in \cite{Gundhi:2018wyz}. This amplification may lead to a peak in the adiabatic power spectrum that is crucial for the formation of PBHs \cite{Gundhi:2020kzm}. 
The power spectra of the scalar perturbations ${\mathcal{R}=HQ_{\sigma}/\dot{\sigma}}$ and ${\mathcal{S}=HQ_{\mathrm{s}}/\dot{\sigma}}$ read
\begin{align}
\mathcal{P}_{\mathcal{R}}={}\frac{k^3}{4\pi^2\varepsilon_{\mathrm{H}}}\frac{\left|Q_{\sigma}\right|^2}{M_{\mathrm{P}}^2},\qquad
\mathcal{P}_{\mathcal{S}}={}\frac{k^3}{4\pi^2\varepsilon_{\mathrm{H}}}\frac{\left|Q_{\mathrm{s}}\right|^2}{M_{\mathrm{P}}^2},
\label{Powerh}
\end{align} 
with the first two Hubble slow-roll parameters defined by
\begin{align}
\varepsilon_{\mathrm{H}}=-\frac{1}{H}\frac{\mathrm{d}\ln H}{\mathrm{d}t},\qquad \eta_{\mathrm{H}}=\frac{1}{H}\frac{\mathrm{d}\ln\varepsilon_{\mathrm{H}}}{\mathrm{d}t}.
\end{align}
\section{Inflation and peak formation}
\label{Sec:InflationAndPeakFormation}
\begin{figure}[!ht]
	\centering
	\begin{tabular}{cc}
		\includegraphics[width=0.45\linewidth]{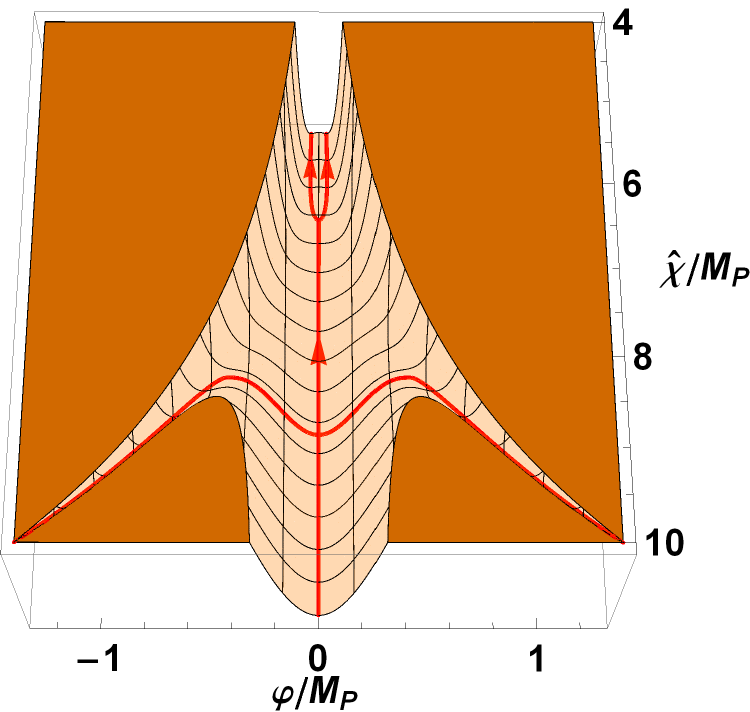}
		\includegraphics[width=0.45\linewidth]{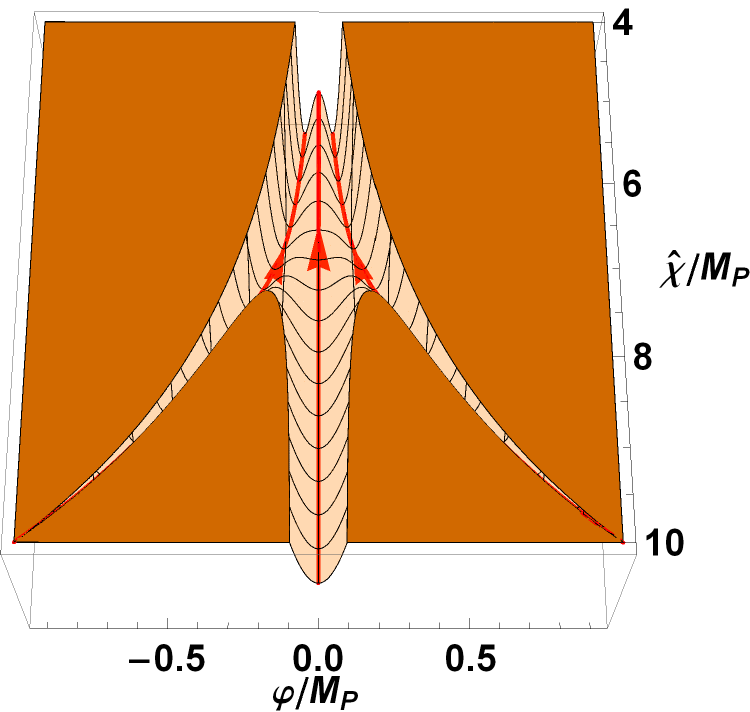}
	\end{tabular}
	\caption{Bird's eye view of the EF two-field potential \eqref{EFpotWExpl}. The red lines sketch the inflationary trajectories running along the valleys in the direction of the arrows. Left: In Scenario I, $\varphi_0$ is a global attractor as the two valleys $\varphi^{\pm}_{\mathrm{v}}$ merge with $\varphi_0$. Right: In Scenario II, the valleys $\varphi_0$ and $\varphi^{\pm}_{\mathrm{v}}$ never merge.}
	\label{PotentialLandscape}
\end{figure}

The landscape of the two-field potential \eqref{EFpotWExpl} is characterized by three valleys at $\varphi_{0}=0$ and $\varphi_{\mathrm{v}}^{\pm}$, which are solutions $\varphi(\hat{\chi})$ of the valley equation
\begin{align}
\hat{W}_{,\varphi}=0.\label{ExtrW}
\end{align}
The model features two different scenarios which are characterized by the parameter combination
\begin{align}
x:={}&6\frac{\xi^2}{\lambda} \frac{m^2_0}{M^2_{\mathrm{P}}}.\label{xdef}
\end{align}
The two scenarios shown in Fig.~\ref{PotentialLandscape} are distinguished by the conditions \cite{Gundhi:2020kzm},
\begin{align}
x<{}&1\qquad\mathrm{Scenario}\; \mathrm{I},\label{S1}\\
x\geq{}&1\qquad\mathrm{Scenario}\; \mathrm{II}.\label{S2}
\end{align}  
We restrict our analysis to \mbox{Scenario I}. In contrast to \mbox{Scenario II}, the global attractor nature of the  ${\varphi_0=0}$ solution in \mbox{Scenario I} ensures that the inflationary background trajectory is independent of the initial conditions.

\noindent In \mbox{Scenario I}, the first stage of inflation proceeds along the $\varphi_0=0$ attractor. It starts at $\hat{\chi}_{\mathrm{i}}$ and lasts until shortly before the critical value $\hat{\chi}_{\mathrm{c}}$ is reached, at which the local $\varphi_0$ minimum turns into an unstable maximum \cite{Gundhi:2020kzm},
\begin{align}\label{ChiCrit}
\hat{\chi}_{\mathrm{c}}=M_{\mathrm{P}}\sqrt{\frac{3}{2}} \ln \left[1+2\frac{\xi }{\zeta }\left(\frac{ m_0 }{M_{\mathrm{P}}}\right)^2\right].
\end{align}
During that stage the inflationary dynamics reduces to an effective single-field model with Starobinsky potential
\begin{align}\label{StarobinskyPotential}
\hat{W}_{\mathrm{Star}}(\hat{\chi}):=\hat{W}(\varphi,\hat{\chi})|_{\varphi=0}=\frac{3}{4}m^2_0M_{\mathrm{P}}^2  \left(1-F^{-1}\right)^2.
\end{align}
Consequently, if $\hat{\chi}_{\mathrm{c}}$ is sufficiently small, the predictions for modes probed by the CMB radiation
\begin{align}
2\times10^{-4}\mathrm{Mpc}^{-1}\lesssim k_{\mathrm{CMB}}\lesssim 2\,\mathrm{Mpc}^{-1},\label{CMBModes}
\end{align}
 are that of Starobinsky's model. For the CMB modes \eqref{CMBModes}, the scalar and tensor power spectra only feature a weak logarithmic $k$ dependence parametrized by the power-law ansatz 
\begin{align}
\mathcal{P}_{h}^{\mathrm{CMB}}\approx{}A_{h}\left(\frac{k}{k_{*}}\right)^{n_{h}},\quad
\mathcal{P}_{\mathcal{R}}^{\mathrm{CMB}}\approx A_{\mathcal{R}}\left(\frac{k}{k_{*}}\right)^{n_{\mathcal{R}}-1}.\label{PowerLawPh}
\end{align} 
Here,  $k_{*}$ is a pivot scale which first crosses the Hubble horizon $N_{*}$ efolds before the end of inflation.
The CMB predictions for $A_{\mathcal{R}}$, $n_{\mathcal{R}}$ and the tensor-to-scalar ratio ${r=A_{h}/A_{\mathcal{R}}}$ in Starobinsky's model read \cite{Starobinsky:1980te},
\begin{align}
A_{\mathcal{R}}^{*}\approx\frac{N_{*}^2}{24 \pi^2 }\frac{m_0^2 }{M_{\mathrm{P}}^2},\quad n_{\mathcal{R}}^{*}\approx &1-\frac{2}{N_{*}},\quad r^{*}\approx\frac{12}{N_{*}^2}.\label{Predictions}
\end{align}
Planck data \cite{Akrami:2018odb} constraints $A_{\mathcal{R}}^{*}$, $n_{\mathcal{R}}^{*}$ at ${k_{*}= 0.05\; \text{Mpc}^{-1}}$,
\begin{align}
A_{\mathcal{R}}^{*}={}&\left(2.099\pm0.014\right)\times 10^{-9}&& (68\%\;\mathrm{CL}),\label{AsCMB}\\
n_{\mathcal{R}}^{*} ={}& 0.9649 \pm 0.0042&& (68\%\;\mathrm{CL}),\label{Planckns}
\end{align}
and the tensor-to-scalar ratio $r^*$ at ${k_{*}= 0.002\; \text{Mpc}^{-1}}$,
\begin{align}
r^{*}<0.064.\label{Planckr}
\end{align}  
For $N_{*}=50\div60$, the predictions \eqref{Predictions} for $n_{\mathcal{R}}^{*}$ and $r^{*}$ are in perfect agreement with \eqref{Planckns} and \eqref{Planckr}.
For $N_{*}=60$, the normalization \eqref{AsCMB} fixes the scalaron mass to be
\begin{align}
m_0\approx1.18\times10^{-5}~M_{\mathrm{P}}=2.8\times 10^{13}~\mathrm{GeV}.\label{scalmass}
\end{align}
According to \eqref{ChiCrit}, by tuning the ratio $\xi/\zeta$ for fixed $m_0^2$, the value of $\hat{\chi}_{\mathrm{c}}$ can be made sufficiently small such that all CMB modes \eqref{CMBModes} cross the horizon before the inflationary trajectory passes the critical point $\hat{\chi}_{\mathrm{c}}$. In this way consistency with the observational constraints \eqref{AsCMB}-\eqref{Planckr} on the spectral CMB observables is ensured.

It is important to note that the relations \eqref{Predictions} only approximately hold in our analysis. This is primarily because the slow-roll dynamics along $\varphi^{\pm}_{\mathrm{v}}$, after crossing the critical point $\hat{\chi}_{\mathrm{c}}$, is faster compared to that in Starobinsky inflation. This implies that  $\hat{\chi}(N_*)>\hat{\chi}_{\mathrm{Star}}(N_*)$, where $\hat{\chi}_{\mathrm{Star}}(N_*)$ represents the numerical value of $\hat{\chi}$ in pure Starobinsky inflation at $N=N_*$. Therefore, in our model, $k_*$ would `feel' a flatter part of the Starobinsky potential at the time of horizon crossing. The spectral index would be closer to one (closer to scale invariance), and therefore slightly higher than that predicted in Starobinsky inflation. We found this deviation to be non-negligible only for the largest values of $\lambda \approx 10 ^{-3}-10^{-2}$ considered in our model. Nevertheless, the spectral index would still be compatible with CMB constrains by choosing $N_*$ to be closer to $50$ efolds for these large values of $\lambda$. In addition, the scalaron mass must be slightly re-adjusted compared to  the value \eqref{scalmass} for different values of $\lambda$. However, as for the spectral index, these numerical changes are insignificant for all parameter values considered in this work. We therefore fix $N_*=60$ and $m_0=1.18\times10^{-5}~M_{\mathrm{P}}$ for all our numerical analyses.

The details of the dynamics in the vicinity of the critical point $\hat{\chi}_{\mathrm{c}}$ are crucial for the formation of a peak in the adiabatic power spectrum. Before reaching $\hat{\chi}_{\mathrm{c}}$, the unit vector $\hat{\sigma}^{I}$ points in $\hat{\chi}$-direction and $\delta\varphi$ is directly related to the isocurvature perturbation $Q_{\mathrm{s}}$ which is suppressed due to a large and positive $m_{\mathrm{s}}^2$, defined in \eqref{IsoMassDef}. At $\hat{\chi}_{\mathrm{c}}$, the effective isocurvature mass $m_{\mathrm{s}}^2$ becomes tachyonic, and, according to \eqref{EQPertQs}, leads to a strong growth of the isocurvature modes $Q_{\mathrm{s}}$. However, already shortly before the critical point is reached, the classical restoring force in $\varphi$-direction (proportional to $m_{\mathrm{s}}^2$) decreases and becomes comparable to the unavoidable zero-point fluctuations $\delta\varphi$ driving the trajectory away from the $\varphi_0$ attractor. During this transition region around $\hat{\chi}_{\mathrm{c}}$ the diffusive quantum effects become important (even dominant) and the inflationary dynamics must be described in terms of a probability density function (PDF) $P(N,\varphi)$ within the stochastic formalism \cite{Starobinsky:1986fx}.\footnote{In  \cite{Pattison:2017mbe,Ezquiaga:2019ftu,Figueroa:2020jkf,Pattison:2021oen} it was found that stochastic effects may induce strong non-Gaussianities relevant for the calculation of the PBH abundance. While non-Gaussianities in general have an impact on the PBH abundance, the intermediate stochastic phase in our model is of a different nature compared to those studied in \cite{Pattison:2017mbe,Ezquiaga:2019ftu,Figueroa:2020jkf,Pattison:2021oen}.}
The PDF gives the probability of the field having the value $\varphi$ at time $N$ and is determined by the Fokker-Planck equation \cite{Gundhi:2020kzm},
\begin{align}\label{FokkerPlanck}
\frac{\partial P}{\partial N}=-\frac{m^2_{\varphi}}{3 H^2} \frac{\partial (\varphi  P)}{\partial \varphi }-\frac{H^2}{8 \pi ^2} \frac{\partial ^2P}{{\partial \varphi} ^2}.
\end{align}  
Here, ${\mathrm{d}N:=-H\mathrm{d}t}$ counts the number of efolds such that ${N=0}$ at the end of inflation. A solution to the Fokker-Planck equation \eqref{FokkerPlanck} is provided by a Gaussian ansatz with time dependent variance $S(N)=\langle\varphi^2\rangle(N)$,
\begin{align}\label{GaussianGuess}
P(\varphi,N)=\frac{1}{\sqrt{2\pi S(N)}}\exp\left(-\frac{\varphi^2}{2S(N)}\right).
\end{align} 
The resulting first-order equation for the variance,
\begin{align}\label{VarianceTimeEv}
\frac{d S}{dN}=\frac{2}{3}\frac{m^2_{\varphi}}{H^2}S -\frac{H^2}{4\pi^2},
\end{align} 
effectively determines the time evolution of the background field $\varphi(N)$ via the identification \cite{Randall:1995dj},
\begin{align}
\varphi(N)\equiv\sqrt{S(N)}.
\end{align}
The stochastic phase during which the quantum diffusive term  $H^2/4\pi^2$ dominates the classical term $2 m^2_{\varphi}S/(3H^2)$ in \eqref{VarianceTimeEv} lasts for a period $\Delta N$ estimated by \cite{Gundhi:2020kzm},
\begin{align}
\Delta N\approx \frac{1}{\zeta}\frac{m^2_0}{M_{\mathrm{P}}^2}.\label{DeltaN}
\end{align}
As discussed in \cite{Gundhi:2020kzm}, for generating a sufficiently sharp peak in the adiabatic power spectrum, the duration $\Delta N$ must be smaller than one efold $\Delta N\lesssim1$.

\begin{figure}[!ht]
	\centering
	\begin{tabular}{cc}
		\includegraphics[width=0.45\linewidth]{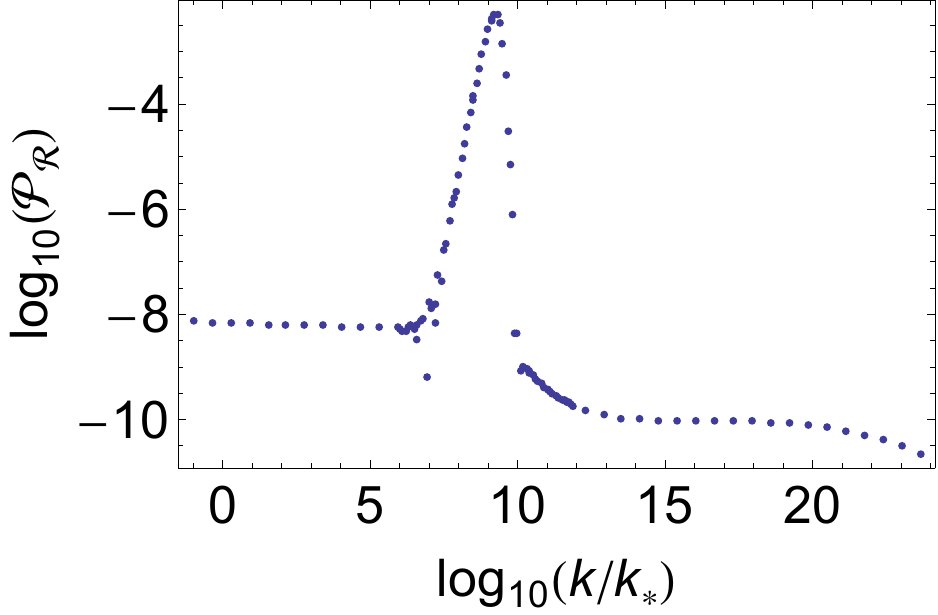}
		\includegraphics[width=0.45\linewidth]{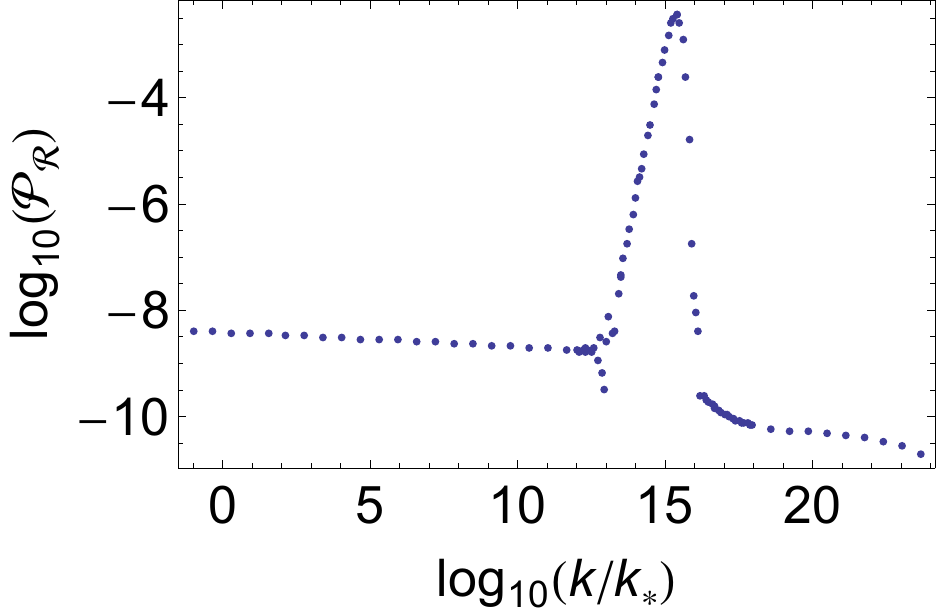}	
	\end{tabular}
	\caption{Log-log plots of the numerically obtained ${\mathcal{P}_{\mathcal{R}}}(k)$ for ${\lambda=10^{-2}}$ with ${k_{*}=0.002\,\mathrm{Mpc}^{-1}}$ and $N_*=60$. Left: The parameters ${\xi=1950}$ and ${\zeta=4.30\times 10^{-9}}$ are chosen such that the power spectrum features a peak at wavenumbers corresponding to the LIGO mass window $M^{III}_{\mathrm{PBH}}$. Right: The parameters $\xi=2050$ and ${\zeta=8.3\times 10^{-9}}$ are chosen such that the power spectrum features a peak at wavenumbers corresponding to the mass window ${M^{II}_{\mathrm{PBH}}}$.}\label{Fig:PowSpectra}
\end{figure}

Once the inflationary trajectory has been driven away from the $\varphi_0$ solution, it turns and falls into one of the $\varphi_{\mathrm{v}}^{\pm}$ valley. According to \eqref{EQPertQsig}, the combination of the non-zero turn rate $\omega$ and the amplified isocurvature modes $Q_{\mathrm{s}}$ leads to a sourcing of the adiabatic modes $Q_{\sigma}$ responsible for the formation of a peak in $\mathcal{P}_{\mathcal{R}}(k)$. The peak formation at the modes $k_{\mathrm{p}}$ corresponding to PBH masses in the two different mass windows $M^{II}_{\mathrm{PBH}}$ and $M^{III}_{\mathrm{PBH}}$ (defined in Sect.~\ref{Sec:PBHAndCDM}) is shown in Fig.~\ref{Fig:PowSpectra} for $\lambda=10^{-2}$.

\noindent After the fall, the background trajectory ultimately settles in one of the outer $\varphi_{\mathrm{v}}^{\pm}$ valleys in which the inflationary dynamics again reduces to an effective single-field model with a second phase of slow-roll inflation. Inflation ends at $\hat{\chi}_{\mathrm{f}}$ determined by the condition $\varepsilon_{\mathrm{H}}(\varphi_{\mathrm{v}}^{\pm},\hat{\chi}_{\mathrm{f}})=1$, close to the global minimum at $\hat{\chi}=0$. The exact inflationary background dynamics is obtained numerically by patching the following three stages as described in detail in \cite{Gundhi:2020kzm}: (\mbox{Stage 1}) Effective single-field Starobinsky inflation along $\varphi_0$ for $\hat{\chi}>\hat{\chi}_{\mathrm{c}}$. (\mbox{Stage 2}) Stochastic phase in the vicinity of $\hat{\chi}_{\mathrm{c}}$. (\mbox{Stage 3}) Fall into the $\varphi_{\mathrm{v}}^{\pm}$ valley and subsequent slow-roll inflation along $\varphi_{\mathrm{v}}^{\pm}$ until $\hat{\chi}_{\mathrm{f}}$.

\section{PBH production and CDM}
\label{Sec:PBHAndCDM}
The generation of a peak in $\mathcal{P}_{\mathcal{R}}(k)$ centered at $k_{\mathrm{p}}$, triggers the production of PBHs with a mass distribution $f(M_{\mathrm{PBH}})$ centered at $M_{\mathrm{PBH}}$ corresponding to $k_{\mathrm{p}}$.  

\begin{figure}[!ht]
	\centering
	\includegraphics[width=0.435\linewidth]{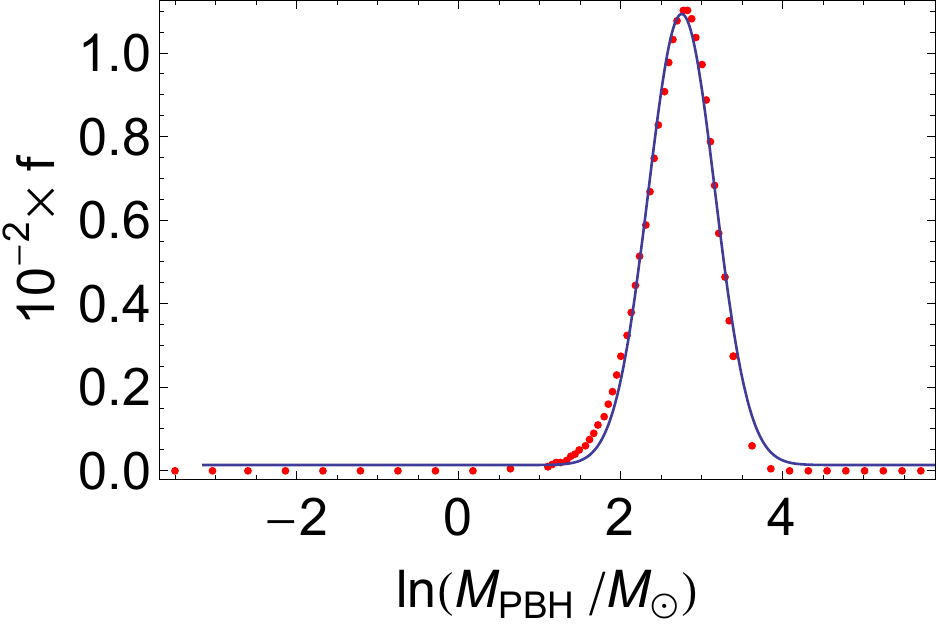} 
	\includegraphics[width=0.435\linewidth]{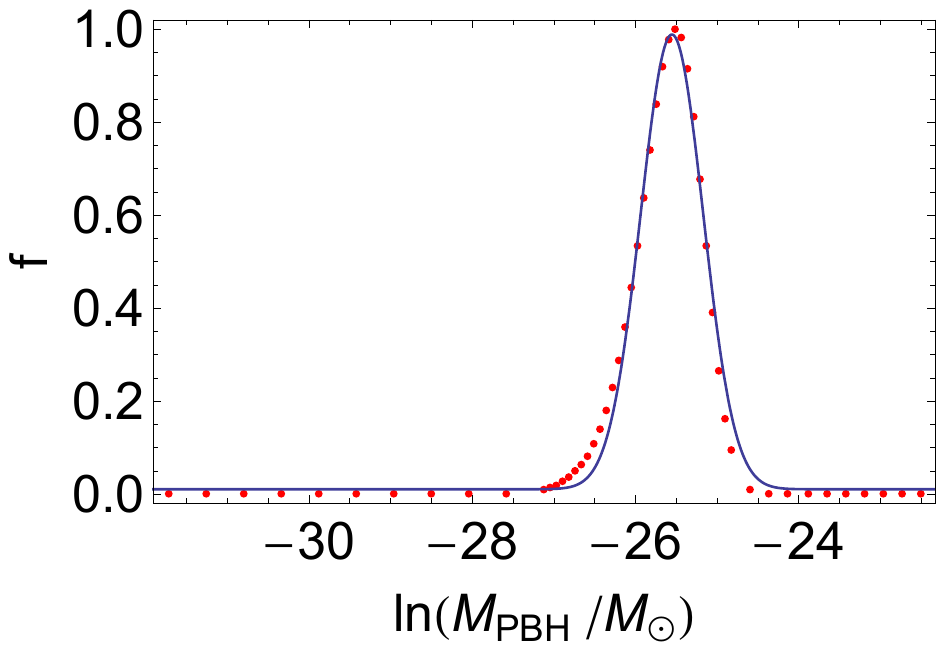}
	\caption{Left: PBH mass distribution $f(M_{\mathrm{PBH}})$ with ${F_{\mathrm{PBH}}=0.01}$ obtained for the LIGO mass window (red dots) with $g(M_{\mathrm{H}})=10.75$ fitted by a log-normal Gaussian \eqref{LogNormal} with ${A_{M}=0.01}$, ${\Delta_M=0.412}$ and ${M_{0}=15.74\,M_{\odot}}$ (blue line). Right: $f(M_{\mathrm{PBH}})$ with ${F_{\mathrm{PBH}}=1}$ obtained for $M^{II}_{\mathrm{PBH}}$ (red dots) with $g(M_{\mathrm{H}})=106.75$ fitted by a log-normal \eqref{LogNormal} with ${A_{M}=0.94}$, ${\Delta_M=0.383}$ and ${M_{0}=8\times 10^{-12}\, M_{\odot}}$ (blue line).}\label{Fig:PBHMassFunction}.
\end{figure}

We assume that a PBH directly forms once an overdensity greater than some critical value $\delta_{\mathrm{c}}$ enters the horizon
\begin{align}
\delta(t,\mathbf{x})=\frac{\rho(t,\mathbf{x})-\bar{\rho}(t)}{\bar{\rho}(t)}.
\end{align}
The PBH mass is given by the critical scaling \cite{Niemeyer:1997mt, Jedamzik:1999am},
\begin{align}\label{CritMPBH}
M_{\mathrm{PBH}}(\delta,t_{\mathrm{f}})=K\,M_{\mathrm{H}}(t_{\mathrm{f}})(\delta-\delta_{\mathrm{c}})^{\gamma}.
\end{align}
Here, $M_{\mathrm{H}}(t_{\mathrm{f}})$ is the horizon mass at the time of formation $t_{\mathrm{f}}$ and the parameters $K$, $\delta_{c}$ and $\gamma$ in \eqref{CritMPBH} are determined numerically \cite{Jedamzik:1999am,Shibata:1999zs,Musco:2004ak,Escriva:2019phb}. As  in \cite{Gundhi:2020kzm}, following the discussion below eq.~(2.16) in \cite{Gow:2020bzo}, we fix these parameters to be $K=10$, $\delta_{\mathrm{c}}=0.25$ and $\gamma=0.36$ consistent with the choice of the window function, as specified in the following discussion.
In the Press-Schechter formalism, the PBH mass fraction $\beta$ at $t_{\mathrm{f}}$ is calculated by \cite{Press:1973iz},
\begin{align}
\beta(t_{\mathrm{f}})=\frac{\rho_{\mathrm{PBH}}(t_{\mathrm{f}})}{\bar{\rho}(t_{\mathrm{f}})}=2\int_{\delta_c}^{\infty}\mathrm{d}\delta\,\frac{M_{\mathrm{PBH}}(\delta,t_{\mathrm{f}})}{M_{\mathrm{H}}(t_{\mathrm{f}})}P(\delta,t_{\mathrm{f}}).\label{beta}
\end{align}  
The Gaussian probability of generating an overdensity with amplitude $\delta$ at $t_{\mathrm{f}}$ is given by
\begin{align}\label{ProbDist}
P(\delta, t_{\mathrm{f}})=\frac{1}{\sqrt{2\pi\sigma^2_{R}(t_{\mathrm{f}})}}\exp{\left(-\frac{1}{2}\frac{\delta^2}{\sigma^2_{R}(t_{\mathrm{f}})}\right)}.
\end{align} 
The variance smoothed over a scale ${R=1/k_{R}}$ with ${k_{R}=a(t_{\mathrm{f}})H(t_{\mathrm{f}})}$ is determined by $\mathcal{P}_{\mathcal{R}}$,
\begin{align}
\sigma_{R}^2(t_{\mathrm{f}})=\int_{0}^{\infty}\mathrm{d}(\ln k)\,\frac{16}{81}\left(\frac{k}{k_R}\right)^4\,W^2(k/k_R)\,\mathcal{P}_{\mathcal{R}}(t_{\mathrm{f}},k).\label{var4}
\end{align}
The Gaussian window function in \eqref{var4} reads \cite{Young:2019osy},
\begin{align}
W(k/k_{R})={}&\exp\left[-\frac{1}{4}\left(\frac{k}{k_R}\right)^2\right].\label{WindowG}
\end{align}
Trading the $t_\mathrm{f}$ dependence for a ${M_{\mathrm{H}}:=M_{\mathrm{H}}(t_{\mathrm{f}})}$ dependence with ${g(t_{\mathrm{eq}}):=g_{\mathrm{eq}}}$ and ${M_{\mathrm{H}}^{\mathrm{eq}}:= M_{\mathrm{H}}(t_{\mathrm{eq}})}$, the PBH mass distribution $f$ as a function of $M_{\mathrm{H}}$ reads \cite{Gundhi:2020kzm},
\begin{align}
f(M_{\mathrm{H}})=\frac{\Omega_{\mathrm{m}}}{\Omega_{\mathrm{c}}}\left(\frac{g(M_{\mathrm{H}})}{g_{\mathrm{eq}}}\right)^{-1/4}\left(\frac{M_{\mathrm{H}}}{M_{\mathrm{H}}^{\mathrm{eq}}}\right)^{-1/2}\beta(M_{\mathrm{H}}).\label{fMh}
\end{align}
We take the matter density parameter ${\Omega_{\mathrm{m}}=0.315}$ and the CDM density parameter $\Omega_{\mathrm{c}}=0.264$ \cite{Aghanim:2018eyx}. 
The total integrated PBH-CDM mass fraction today is defined as 
\begin{align}\label{Ftotal}
F_{\mathrm{PBH}}:=\int_{-\infty}^{\infty}f(M_{\mathrm{H}})\mathrm{d}\ln M_{\mathrm{H}}.
\end{align}
Following the analysis in \cite{Gundhi:2020kzm}, the mass distribution as a function of the PBH mass $M_{\mathrm{PBH}}$ is defined by ${F_{\mathrm{PBH}}:=\int f(M_{\mathrm{PBH}})\mathrm{d}\ln M_{\mathrm{PBH}}} $ and obtained by using \eqref{CritMPBH}, \eqref{beta}, \eqref{ProbDist}, \eqref{fMh}, and \eqref{Ftotal} with $\mu:= M_{\mathrm{PBH}}/(K M_{\mathrm{H}})$,
\begin{align}
f(M_{\mathrm{PBH}})=2\frac{\Omega_{\mathrm{m}}}{\Omega_{\mathrm{c}}}{}&\int_{-\infty}^{\infty} d(\ln M_{\mathrm{H}})\frac{M_{\mathrm{PBH}}}{M_{\mathrm{H}}}\left(\frac{g(M_{\mathrm{H}})}{g_{\mathrm{eq}}}\right)^{-1/4}\nonumber\\
&\times\left(\frac{M_{\mathrm{H}}}{M^{\mathrm{eq}}_{\mathrm{H}}}\right)^{-1/2}\frac{\mu^{1/\gamma}}{\gamma\sqrt{2\pi\sigma_{R}^2(M_{\mathrm{H}})}}\nonumber\\
&\times\exp\left(-\frac{\left(\mu^{1/\gamma}+\delta_{\mathrm{c}}\right)^2}{2\sigma_{R}^2(M_{\mathrm{H}})}\right).\label{fpbhfinal}
\end{align}
A simple estimate for the approximate PBH mass as a function of the peak scale $k_{\mathrm{p}}$ is obtained as \cite{Nakama:2016gzw},
\begin{align}
M_{\mathrm{PBH}}(k_{\mathrm{p}})\approx 6.3\times 10^{12}M_{\odot} \left(\frac{k_{\mathrm{p}}}{\mathrm{Mpc}^{-1}}\right)^{-2}.\label{MPBHEst}
\end{align}	
The PBH mass windows for which $F_{\mathrm{PBH}}=1$ is compatible with observational constraints are \cite{Carr:2020gox, Carr:2020xqk},
\begin{align}
10^{-17}M_{\odot}\lesssim{}& M_{\mathrm{PBH}}^{I}\lesssim 10^{-16}M_{\odot}\label{M1},\\
10^{-13}M_{\odot}\lesssim{}& M_{\mathrm{PBH}}^{II}\lesssim 10^{-9}M_{\odot}\label{M2}.
\end{align}
There is an ongoing discussion about the possibility to explain all the observed CDM by PBHs in the LIGO mass window $M_{\mathrm{PBH}}^{III}$  \cite{Carr:2020gox, Carr:2020xqk, Sasaki:2018dmp},
\begin{align}
10 M_{\odot}\lesssim M_{\mathrm{PBH}}^{III}\lesssim 10^2 M_{\odot}.\label{M3}
\end{align}
According to \cite{Carr:2020xqk,Chen:2018czv,Raidal:2018bbj,Gow:2019pok,DeLuca:2020qqa}, a maximal contribution ${F_{\mathrm{PBH}}\lesssim 10^{-2}\div10^{-3}}$ in $M_{\mathrm{PBH}}^{III}$ seems to be favored. 
The relation \eqref{MPBHEst} implies that the mass windows \eqref{M1}-\eqref{M3} are related to the following peak scales $k_{\mathrm{p}}$,
\begin{align}
k^{I}_{\mathrm{p}}\approx{}& 10^{15}\mathrm{Mpc}^{-1},\\
10^{13}\mathrm{Mpc}^{-1}\gtrsim{}& k^{II}_{\mathrm{p}}\gtrsim 10^{11}\mathrm{Mpc}^{-1},\\
k^{III}_{\mathrm{p}}\approx{}& 10^{6}\mathrm{Mpc}^{-1}.
\end{align}
We present the PBH mass distribution $f(M_{\mathrm{PBH}})$ for the LIGO mass window $M_{\mathrm{PBH}}^{III}$ and the mass window $M^{II}_{\mathrm{PBH}}$ in Fig.~\ref{Fig:PBHMassFunction}.

Both mass distributions are obtained for the inflationary power spectra shown in Fig.~\ref{Fig:PowSpectra} and are consistent with all observational constraints. The numerically generated $f(M_{\mathrm{PBH}})$ (red dots) are fitted well by a log-normal distribution (blue line) defined by
\begin{align}
f(M_{\mathrm{PBH}})=\frac{A_{M}}{\sqrt{2\pi\Delta_M^2}}\exp\left\{-\frac{\left[\ln \left(M_{\mathrm{PBH}}/M_{0}\right)\right]^2}{2\Delta_M^2}\right\}\label{LogNormal}.
\end{align}

For completeness, we mention that there are also parameter combinations leading to a mass distribution $f(M_{\mathrm{PBH}})$, which is compatible with observational constraints and leads to $F_{\mathrm{PBH}}=1$ in the mass window $M_{\mathrm{PBH}}^{I}$. Since recent data from the NANOGrav Collaboration \cite{Arzoumanian:2020vkk} suggests that PBHs may constitute a large part (if not all) of CDM with a mass distribution centered in the mass window  $10^{-15}M_{\odot}\div 10^{-11} M_{\odot}$ \cite{DeLuca:2020agl}, the distinction between the mass windows $M_{\mathrm{PBH}}^{I}$ and $M_{\mathrm{PBH}}^{II}$ might become obsolete.

\section{Constraints on parameters}
We have demonstrated that there are parameter combinations for which the extended scalaron-Higgs model describes a successful phase of inflation yielding predictions for the spectral observables \eqref{Predictions} that are in perfect agreement with Planck data \eqref{AsCMB}-\eqref{Planckr}. At the same time, it explains the observed CDM in terms of PBHs.
The compatibility with CMB measurements requires fixing $m_0$ according to \eqref{scalmass}. The quartic Higgs coupling $\lambda$ is determined by SM physics. The remaining free parameters $\zeta$ and $\xi$ are fixed by the properties of the peak in $\mathcal{P}_{\mathcal{R}}(k)$ which ultimately determines the PBH fraction of the observed CDM. 
In this section, we derive general scaling relations among the parameters $\lambda$, $\zeta$, and $\xi$, for a given $F_{\mathrm{PBH}}$ with underlying mass distribution $f(M_{\mathrm{PBH}})$ centered at a given $M_{\mathrm{PBH}}$.  

First, we derive a direct relation between the model parameters $\zeta$ and $\xi$ and the PBH mass around which $f(M_{\mathrm{PBH}})$ is centered.
Using the definition of the number of efolds $N_*-N=\ln a/a_*$, for modes which cross the horizon at $k=aH$ and $k_*=a_*H$ with constant $H\approx H_*$ respectively, we obtain $N_*-N=\ln k/k_*$. During the phase of effective Starobinsky inflation, there is a simple relation between $N$  and the field value $\hat{\chi}$ given by $N(\hat{\chi})\approx F(\hat{\chi})$. The peak in $\mathcal{P}_{\mathcal{R}}(k)$ is centered around the modes $k\approx k_{\mathrm{p}}\pm\Delta k$ which cross the horizon in the vicinity of $\hat{\chi}_{\mathrm{c}}$. Hence, we can express $k_{\mathrm{p}}$ in terms of $\hat{\chi}_{\mathrm{c}}$ defined in \eqref{ChiCrit} by the relation
\begin{align}
k_{\mathrm{p}}\approx k_*\exp\left(N_{*}-1-2\frac{\xi}{\zeta}\frac{m_0^2}{M_{\mathrm{P}}^2}\right).\label{kpPara}
\end{align}
Since $M_{\mathrm{PBH}}$ is related to the peak scale $k_{\mathrm{p}}$ via \eqref{MPBHEst} we finally obtain $M_{\mathrm{PBH}}$ in terms of the model parameters, the pivot scale $k_*$ and the total number of efolds $N_*$,
\begin{align}
M_{\mathrm{PBH}}\approx M_{\odot} \left(\frac{4\times 10^{-7}\,k_{*}}{\mathrm{Mpc}^{-1}}\right)^{-2}e^{-2(N_{*}-1)+4\frac{\xi}{\zeta}\frac{m_0^2}{M_{\mathrm{P}}^2}}.
\end{align} 
For $N_*=60$, $k_*=0.002\,\mathrm{Mpc}^{-1}$, and $m_0$ as in \eqref{scalmass}, we obtain the linear scaling relation between $\zeta$ and $\xi$ with a  $M_{\mathrm{PBH}}$-dependent proportionality coefficient
\begin{align}
\zeta\approx 5.6\times 10^{-10}\left[\ln\left(10^{33}\frac{M_{\mathrm{PBH}}}{M_{\odot}}\right)\right]^{-1}\,\xi.\label{LinScaling}
\end{align}

Next, we obtain a relation involving $\lambda$ and $\xi$ from the requirement that $F_{\mathrm{PBH}}$ acquires a specific value for a mass distribution $f(M_{\mathrm{PBH}})$ centered around a given $M_{\mathrm{PBH}}$. In view of the complex inflationary dynamics around peak formation, described in Sect.~\ref{Sec:InflationAndPeakFormation}, and the details involved in calculating $F_{\mathrm{PBH}}$ described in Sect.~\ref{Sec:PBHAndCDM}, going beyond an order of magnitude estimate based on various simplifying assumptions seems to be illusive.

We start by noting that $\mathcal{P}_{\mathcal{R}}(k)$ is related to the perturbation $Q_{\sigma}(N,\mathbf{x})$ in position space by 
\begin{align}
\int d\ln k\mathcal{P}_{\mathcal{R}}(k)=\frac{1}{2\varepsilon _{\mathrm{H}}}\frac{\left\langle Q^2_{\sigma}(N,\mathbf{x})\right\rangle}{M^2_{\mathrm{P}}}.\label{PS}
\end{align}
According to \eqref{QSigma}, $Q_{\sigma}$ is related to $\delta\hat{\chi}$ and $\delta\varphi$ via ${Q_{\sigma}=G_{IJ}\hat{\sigma}^I\delta\Phi^J}$, with $\Phi^{I}$ and $G_{IJ}$ defined in \eqref{GMetric}. During most of the inflationary dynamics along $\varphi_0$ and the later part of the dynamics in $\varphi_{\mathrm{v}}^{\pm}$, the inflaton vector $\hat{\sigma}^{I}$ points in the $\hat{\chi}$ direction and $Q_{\sigma}$ exclusively receives contribution from $\delta\hat{\chi}$. Only during the short peak formation stage in the vicinity of $\hat{\chi}_{\mathrm{c}}$, where the trajectory turns and $\hat{\sigma}^I$ has a non-zero component in $\varphi$-direction, $Q_{\sigma}$ also receives contribution from $\delta\varphi$. Here, we assume that during this period $\hat{\sigma}^{I}$ points in the $\varphi$-direction such that $(\hat{\sigma}^{\varphi})^2=F$.\footnote{The exact dynamics is more complicated and involves a short phase in which $\delta\varphi$ and $\delta\hat{\chi}$ simultaneously contribute to $Q_{\sigma}$. 
}  
For modes $k\approx k_{\mathrm{p}}\pm\Delta k$, which cross the horizon during this period, \eqref{PS} reduces to
\begin{align}
\int_{k_{\mathrm{p}}-\Delta k}^{k_{\mathrm{p}}+\Delta k} d\ln k\mathcal{P}_{\mathcal{R}}(k) \approx
\frac{1}{2\varepsilon_{\mathrm{H}}F}\frac{\langle\delta\varphi^2\rangle}{M^2_{\mathrm{P}}}.\label{IntDk}
\end{align}
For a simplified treatment we take the sharp peak limit $\mathcal{P}_{\mathcal{R}}(k)\approx A_{\mathrm{p}}\delta(\ln k-\ln k_{\mathrm{p}})$ such that \eqref{IntDk} becomes
\begin{align}
A_{\mathrm{p}}\approx \frac{1}{2\varepsilon_{\mathrm{H}}F}\frac{\langle\delta\varphi^2\rangle}{M^2_{\mathrm{P}}}.\label{ApApprox}
\end{align}
Although the slow-roll dynamics along $\varphi^{\pm}_{\mathrm{v}}$ slightly differs from that of the effective Starobinsky inflation along $\varphi_0$, for an order of magnitude estimate we use the background relations of Starobinsky inflation ${F(\hat{\chi})\approx N}$ and ${\varepsilon_{\mathrm{H}}(N)\approx 1/N^2}$ evaluated at $N_{\mathrm{c}}:=N(\hat{\chi}_{\mathrm{c}})$, so that \eqref{ApApprox} reduces to\footnote{To produce PBHs in the mass windows $M_{\mathrm{PBH}}^{III}$ and $M_{\mathrm{PBH}}^{II}$, we find $N_{\mathrm{c}}\approx40$ and $N_{\mathrm{c}}\approx 25$, respectively.}
\begin{align}
A_{\mathrm{p}}\approx\frac{N_{\mathrm{c}}}{2}\frac{\langle\delta\varphi^2\rangle}{M^2_{\mathrm{P}}}.\label{peakamplapprox}
\end{align}
As discussed in Sect.~\ref{Sec:InflationAndPeakFormation}, close to $\hat{\chi}_{\mathrm{c}}$ quantum diffusive effects dominate and a stochastic treatment is required during which $\varphi(N)$ is identified with $\langle\delta\varphi^2(N,\mathrm{x})\rangle^{1/2}$. But even after the stochastic phase, during the fall from $\varphi_0$ to $\varphi_{\mathrm{v}}^{\pm}$, both $\delta\varphi$ and $\varphi$ continue to grow together -- $\delta\varphi$ because $\hat{W}_{,\varphi\varphi}$ is still negative, and $\varphi$ because it moves away from $\varphi=0$ to larger field values until it reaches $\varphi_{\mathrm{v}}^{\pm}$. However, just before the background trajectory settles in the $\varphi_{\mathrm{v}}^{\pm}$ valley,  $\hat{W}_{,\varphi\varphi}$ turns positive at the inflection point $\hat{W}_{,\varphi\varphi}=0$ during the fall. This leads to a sudden stop of the growth of $\delta\varphi$, while $\varphi$ still continues to grow until $\varphi=\varphi_{\mathrm{v}}^{\pm}$. Hence $\sqrt{\langle \delta\varphi^2\rangle}$ is bounded from above by the maximum distance between the two valleys $\varphi_{\mathrm{v}}^{\pm}(\hat{\chi}_{\mathrm{max}})-\varphi_0=\varphi_{\mathrm{v}}^{\pm}(\hat{\chi}_{\mathrm{max}})$ attained at $\hat{\chi}_{\mathrm{max}}<\hat{\chi}_{\mathrm{c}}$,
\begin{align}
\langle\delta\varphi^2\rangle \leq \left|\varphi_{\mathrm{v}}^{\pm}(\hat{\chi}_{\mathrm{max}})\right|^2, \qquad \left.\frac{\partial\varphi^{\pm}_{\mathrm{v}}(\hat{\chi})}{\partial \hat{\chi}}\right|_{\hat{\chi}_{\mathrm{max}}}=0.\label{Estimate}
\end{align} 
A strong amplification of $\mathcal{P}_{\mathcal{R}}(k)$ requires a large $\delta\varphi$ and hence a large $|\varphi^{\pm}_{\mathrm{v}}(\hat{\chi}_{\mathrm{max}})|$.
The inequality in \eqref{Estimate} can be parametrized by $\langle\delta\varphi^2\rangle\approx \alpha^2|\varphi^{\pm}_{\mathrm{v}}(\hat{\chi}_{\mathrm{max}})|^2$ with ${\alpha\in[0.1,1]}$.\footnote{Geometrically, the inflection point which lies between $\varphi_0$ and $\varphi_{\mathrm{v}}^{\pm}$ cannot be too close to $\varphi_0=0$. In addition, the inertia of the background dynamics carries the trajectory along $\varphi_0$ even after reaching the bifurcation point shown in the left plot of Fig.~\ref{PotentialLandscape}, such that the fall into $\varphi^{\pm}_{\mathrm{v}}$ happens only after the valleys reach a sufficient separation, justifying the lower bound on $\alpha$. }
The analytic expression for $|\varphi^{\pm}_{\mathrm{v}}(\hat{\chi}_{\mathrm{max}})|^2$ is found from \eqref{ExtrW} and the second equation in \eqref{Estimate} as\footnote{The criterion to determine $\hat{\chi}_{\mathrm{max}}$ in \eqref{Estimate} only applies to \mbox{Scenario I}. Only in this scenario, the valleys re-emerge at the bifurcation point $\hat{\chi}_{\mathrm{c}}$ turn and again move towards $\varphi=0$.}
\begin{align}\label{PhiMaxExact}
|\varphi^{\pm}_{\mathrm{v}}(\hat{\chi}_{\mathrm{max}})|^2=\frac{m_0^2}{\zeta} L(x),
\end{align}
with $x$ defined in \eqref{xdef} and the function $L(x)$ defined by
\begin{align}
L(x):=\frac{2-2\sqrt{1-x}-x}{x}.\label{Lfunc}
\end{align}
Since \eqref{S1} implies $x<1$, the function \eqref{Lfunc} takes arguments from $x\in[0,1)$ which means that \eqref{Lfunc} takes values in the interval ${L(x)\in[0,1)}$. For an order of magnitude estimate we approximate ${L(x)=x/4+\mathcal{O}(x^2)}$ and obtain
\begin{align}
\langle\delta\varphi^2\rangle\approx\alpha^2|\varphi^{\pm}_{\mathrm{v}}(\hat{\chi}_{\mathrm{max}})|^2\approx\frac{\alpha^2m^2_0}{4\zeta}x.\label{dphiav}
\end{align}
Combining \eqref{peakamplapprox} with \eqref{dphiav}, we obtain the analytic estimate for the peak amplitude
\begin{align}\label{AnalyticEstimate}
A_{\mathrm{p}}\approx\frac{N_{\mathrm{c}}}{8\zeta }\frac{m^2_0}{M^2_{\mathrm{P}}}\alpha^2 x.
\end{align}
We argued in Sect.~\ref{Sec:InflationAndPeakFormation} that $\Delta N\approx m^2_0/(M^2_{\mathrm{P}}\zeta)$ corresponds to the duration of the stochastic phase and that this phase must be sufficiently short $\Delta N\lesssim1$ in order to produce a narrow peak in $\mathcal{P}_{\mathcal{R}}(k)$. Since $N_{\mathrm{c}}=\mathcal{O}(10)$, the magnitude of the total prefactor in \eqref{AnalyticEstimate} is estimated to be of order $N_{\mathrm{c}}\,\Delta N \alpha^2/8\approx 10^{-2}$, leading to the condition
\begin{align}\label{AnalyticEstimate2}
A_{\mathrm{p}}\approx 10^{-2}x.
\end{align}
Since a significant $F_{\mathrm{PBH}}\approx1$ requires a peak amplitude $A_{\mathrm{p}}\approx 10^{-2}\div10^{-3}$ \cite{Sasaki:2018dmp,Gundhi:2020kzm}, it is clear that $x$ cannot be much smaller than one and we finally obtain the estimate
\begin{align}
x\approx 1.
\end{align}
Using \eqref{xdef}, this yields the approximate scaling relation 
\begin{align}\label{XiConstraint}
\lambda\approx 6\frac{m^2_0}{M^2_{\mathrm{P}}}\xi^2.
\end{align}
Inserting the scalaron mass $m_0$ from \eqref{scalmass}, we obtain 
\begin{align}\label{XiConstraintNum}
\lambda\approx 10^{-9}\xi^2.
\end{align}
The precise value of $F_{\mathrm{PBH}}$ for a given power spectrum $\mathcal{P}_{\mathcal{R}}(k)\approx A_{\mathrm{p}}\delta(\ln k-\ln k_{\mathrm{p}})$ is exponentially sensitive to the peak amplitude $A_{\mathrm{p}}$, as can be seen from the relations \eqref{beta}-\eqref{Ftotal}. This is the main reason why any attempt to obtain a precise analytical relation for $F_{\mathrm{PBH}}$ in terms of the model parameters is hard to realize. 
Nevertheless, in view of \eqref{AnalyticEstimate2}, the amplification only depends on $x$ such that the same amplification is achieved for different values of $\xi$ and $\lambda$ as long as they are related by the scaling relation \eqref{XiConstraintNum}. In general the exact numerical factor in the quadratic scaling law \eqref{XiConstraintNum}, depends on the values of $F_{\mathrm{PBH}}$ and the PBH mass $M_{\mathrm{PBH}}$ at which the mass distribution $f(M_{\mathrm{PBH}})$ peaks, but the scaling law $\lambda\propto\xi^2$ will be the same for all mass windows and total mass fractions.

As a side note, the relation \eqref{XiConstraintNum} coincides with the CMB normalization condition found in pure Higgs inflation \cite{Bezrukov2008,Barvinsky2008,Bezrukov2009a,DeSimone2009,Barvinsky2009}. This coincidence is surprising, as in our model the CMB normalization condition \eqref{AsCMB} is satisfied by $m_0$ alone and the parameters $\lambda$ and $\xi$ are not directly related to CMB physics at large wavelengths but rather determine the PBH formation resulting from a peak in $\mathcal{P}_{\mathrm{R}}$ at small wavelengths. 

\begin{figure}[!ht]
	\centering
	\includegraphics[width=0.45\linewidth]{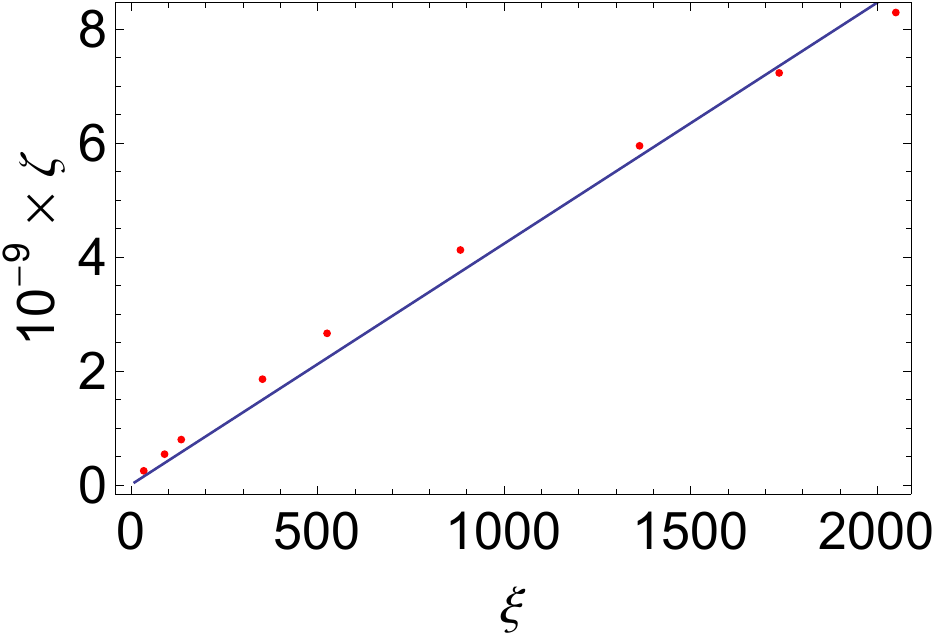}
	\includegraphics[width=0.49\linewidth]{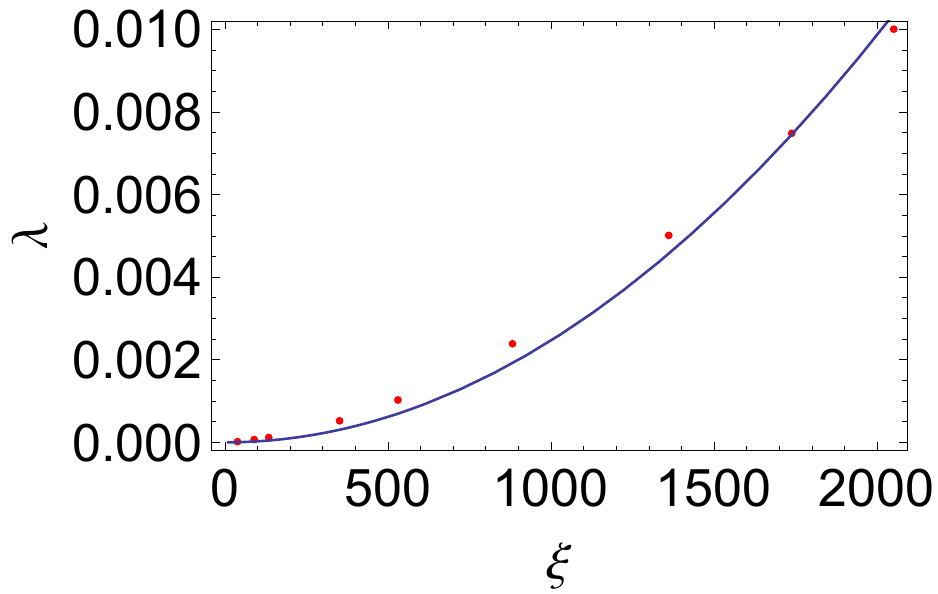}
	\caption{Numerically obtained scaling relations for the parameters leading to $F_{\mathrm{PBH}}\approx1$ for $f(M_{\mathrm{PBH}})$ centered around $M_{\mathrm{PBH}}=10^{-11}M_{\odot}$. Left: Linear scaling relation between $\zeta$ and $\xi$. Numerically generated points (red) linear fit (blue). Right: Quadratic scaling relation between $\lambda$ and $\xi$. Numerically generated points (red) quadratic fit (blue). }\label{Fig:LambdaXi}
\end{figure}

Finally, we check the analytical estimates \eqref{LinScaling} and \eqref{XiConstraint} by an exact numerical analysis. We systematically perform a parameter scan for different values of $\lambda$, $\xi$ and $\zeta$ such that a mass distribution $f(M_{\mathrm{PBH}})$ in the window $M^{II}_{\mathrm{PBH}}$ centered around $M_{\mathrm{PBH}}=10^{-11}M_{\odot}$ with $F_{\mathrm{PBH}}\approx1$, as shown in the right plot of Fig.~\ref{Fig:PBHMassFunction}, is realized. The parameters $\lambda$, $\xi$ and $\zeta$  that permit such realizations, are related to each other by the scaling relations shown in Fig.~\ref{Fig:LambdaXi}, which are remarkably close to the analytical estimates \eqref{LinScaling} and \eqref{XiConstraint}.

\noindent The linear and quadratic fits to the numerically found scaling relations in Fig.~\ref{Fig:LambdaXi} are given by
\begin{align}
 \zeta={}& 4.23\times 10^{-12}\xi,\qquad \lambda = 2.47\times 10^{-9}\xi^2.\label{fits}
\end{align}
In addition to the correct functional form of the scaling relations, also the numerical coefficients in \eqref{fits} agree well with those predicted by the analytic estimates \eqref{LinScaling} and \eqref{XiConstraintNum}, thereby numerically confirming them.

All parameters of the extended scalaron-Higgs model are fixed. 
The parameter $m_0$ is fixed by the CMB constraint \eqref{scalmass} on the scalar inflationary power spectra at large wavelengths, independently of the value for the quartic Higgs coupling $\lambda$. In contrast, the non-minimal couplings $\zeta$ and $\xi$ are ultimately determined in terms of $\lambda$ by the scaling relations \eqref{LinScaling} and \eqref{XiConstraint}. The relations are determined by the requirement that the peak in $\mathcal{P}_{\mathcal{R}}(k)$ leads to a significant $F_{\mathrm{PBH}}$ with a PBH mass distribution $f(M_{\mathrm{PBH}})$ centered around $M_{\mathrm{PBH}}$.

In the SM, the tree-level value of the quartic Higgs coupling  $\lambda\approx 10^{-1}$ is determined by the symmetry breaking scale $\nu$ and the Higgs mass $M_{\mathrm{h}}$. In view of the huge energy gap separating the electroweak energy scale and the inflationary energy scale, the RG improvement becomes crucial to determine the value of the running Higgs coupling $\lambda$ during inflation \cite{Bezrukov2009a,DeSimone2009,Barvinsky2009}.
The contributions to the beta function of $\lambda$ are dominated by quantum loops of the heavy SM particles.
The system of the RG equations is highly sensitive to the precise conditions of the RG flow at the electroweak scale, in particular to the Higgs mass $M_{\mathrm{h}}$ and the Yukawa top-quark mass, which are constraint by recent experimental bounds \cite{Tanabashi2018},
\begin{align}
M_{\mathrm{h}}={}&125.10\pm 0.14\;\mathrm{GeV},\label{Mh}\\
M_{\mathrm{t}}={}&172.9\pm 0.4\;\mathrm{GeV}.\label{Mt}
\end{align}
While an analysis of the full RG system of the extended scalaron-Higgs model would be required for a precise determination of the running $\lambda$ at the inflationary energy scale, already the pure SM running may be sufficient to derive a lower bound on $\lambda$.
The RG running based on the SM beta functions drives $\lambda$ to small values at high energies during inflation. Depending on \eqref{Mh} and \eqref{Mt}, $\lambda$ might even become negative and trigger an instability of the RG improved effective Higgs potential \cite{Degrassi2012,Buttazzo2013}. Assuming a stable $\lambda>0$, by continuously varying the masses \eqref{Mh} and \eqref{Mt}, the value of $\lambda$ can in principle be made arbitrarily small at some energy scale. However, the smallest value of $\lambda$ which can be sustained over the course of the inflationary dynamics is found to be $\lambda\approx 10^{-6}$  \cite{Hamada2014,Bezrukov2014,Hamada2015,Gundhi:2018wyz}. 

Hence, once the total mass fraction $F_{\mathrm{PBH}}\leq1$ and the PBH mass $M_{\mathrm{PBH}}$ around which the mass distribution $f(M_{\mathrm{PBH}})$ is centered are specified, the values of $\zeta$ and $\xi$ are fixed in terms of $\lambda$ via the scaling relations \eqref{LinScaling} and \eqref{XiConstraintNum}. The value of $\lambda=10^{-2}\div10^{-6}$ during inflation, in turn, is fixed by SM physics at the electroweak scale. In this way, our unified model incorporates the physics of the SM at the electroweak scale, explains the presently observed CDM content of the Universe by PBHs and leads to inflationary predictions in agreement with measurements of the CMB radiation.

\section{Conclusions}
The extended scalaron-Higgs model proposed in this article is a viable model of inflation, which, at the same time, explains the origin of the presently observed CDM by PBHs. 

One of the main features of the model is a Higgs-dependent scalaron mass which arises from a non-minimal coupling of the SM Higgs field to the quadratic scalar curvature invariant. Compared to the scalaron-Higgs model \cite{Ema2017a,Wang2017,He:2018gyf,Gundhi:2018wyz}, the additional non-minimal coupling $\zeta$ introduces one more parameter.
With this additional parameter, the physics of the early Universe and the physics of the SM at the electroweak scale are described in one unified model which explains the observed CDM content without assuming any new particle, except for the scalaron which effectively emerges from the modified gravitational sector. 

In addition, due to the global attractor nature of the $\varphi_0$ solution, the scenario considered in this article has the appealing feature that its predictions do not depend on the initial conditions of the inflationary background trajectory.
A correct description of the background dynamics requires a stochastic treatment in the vicinity of the critical point $\hat{\chi}_{\mathrm{c}}$.
The inclusion of these diffusive quantum effects are crucial for an accurate quantitative treatment of the multi-field ``isocurvature pumping'' mechanism and leads to the formation of a peak in $\mathcal{P}_{\mathcal{R}}$ at small wavelengths responsible for a significant production of PBHs \cite{Gundhi:2020kzm} .

We find that the extended scalaron-Higgs model can produce an observationally viable mass distribution $f(M_{\mathrm{PBH}})$ with $F_{\mathrm{PBH}}\approx10^{-2}\div10^{-3}$ in the LIGO mass window \eqref{M3} and $F_{\mathrm{PBH}}\approx1$ in the mass windows \eqref{M1} and \eqref{M2}.
We find simple scaling relations \eqref{LinScaling} and \eqref{XiConstraint} between the non-minimal coupling parameters $\zeta$, $\xi$, and $\lambda$, for CDM to be described by PBHs of a given mass. Together with the CMB normalization condition \eqref{AsCMB}, which fixes the scalaron mass $m_0$ via \eqref{Predictions}, these scaling relations uniquely determine all parameters of the model in terms of $\lambda$. The quartic Higgs coupling $\lambda$, in turn, is determined by SM physics. The RG analysis of the SM suggests that a positive $\lambda$ can take values $\lambda\approx 10^{-2}\div10^{-6}$ at the energy scale of inflation.
For these values of $\lambda$, the model permits a viable phase of inflation and an explanation of the observed CDM through PBH production in one single unified model without assuming any new physics.  

The predictions of the model on wavelengths probed by the CMB are identical to that of Starobinsky's model. Thus, a measurement of the tensor-to-scalar ratio higher than that predicted by Starobinsky's model would rule out this model. 

Another characteristic feature of the model is that it can only produce a single peak in the adiabatic power spectrum, such that $F_{\mathrm{PBH}}$ can only receive contributions from PBHs within a narrow mass interval.
Therefore, the model can also be tested against the possibility of $F_{\mathrm{PBH}}$ collecting significant contributions from PBHs in different mass intervals. 

Finally, since all parameters of the model are fixed by the SM ($\lambda$), the CMB ($m_0$), and a significant $F_{\mathrm{PBH}}$ ($\xi$) for $f(M_{\mathrm{PBH}})$ centered at $M_{\mathrm{PBH}}$ ($\zeta$), any additional prediction of the model, such as the production of gravitational waves accompanying the formation of PBHs \cite{Nakama:2016gzw,Garcia-Bellido:2017aan,Sasaki:2018dmp,Belotsky:2018wph}, which may be detected by the space-based gravitational interferometer LISA \cite{Cai:2018dig,Bartolo:2018evs,Braglia:2020eai}, could potentially rule out this model.

\section*{Acknowledgments}
AG and CFS thank Sergey V.~Ketov for discussions.
AG thanks the University of Trieste and INFN for
financial support and Angelo Bassi for supporting this
collaboration. 
\bibliography{PBHScalaronHiggsBib}{}
\end{document}